\documentclass{ws-procs975x65}

\begin{document}

\title{Precision timing of PSR J1012+5307 and strong-field GR tests}

\author{KOSMAS LAZARIDIS$^*$, NORBERT WEX, AXEL JESSNER, MICHAEL KRAMER$^{**}$, \\ J. ANTON ZENSUS}

\address{Max-Planck-Institut f\"ur Radioastronomie\\
Auf dem H\"ugel 69, Bonn, 53121, Germany\\
$^*$E-mail: klazarid@mpifr-bonn.mpg.de}

\author{BEN W. STAPPERS$^{***}$, GEMMA H. JANSSEN, MARK B. PURVER, ANDREW G. LYNE, CHRISTINE A. JORDAN}

\address{$^{**}$Jodrell Bank Centre for Astrophysics, Alan Turing Building, \\
School of Physics and Astronomy,\\
 University of Manchester, Manchester, M13 9PL, UK}

\address{$^{***}$Stichting ASTRON, \\ Dwingeloo, Postbus 2, 7990 AA, the
  Netherlands}

\author{GREGORY DESVIGNES, ISMAEL COGNARD, GILLES THEUREAU}

\address{Laboratoire de Physique et Chimie de l'Environnement, \\ CNRS, 3A
  Avenue de la Recherche Scientifique, \\ Orl\'{e}ans Cedex
  2, 45071,  France \\}

\address{Station de Radioastronomie de Nan\c cay, Paris Observatory, \\
  University of Orl\'{e}ans, CNRS/INSU, 18330, Nan\c cay, France}

\begin{abstract}
We report on the high precision timing analysis of the pulsar-white dwarf binary PSR J1012+5307. Using 15 years of multi-telescope data from the European Pulsar Timing Array (EPTA) network, a significant measurement of the variation of the orbital period is obtained. Using this ideal strong-field gravity laboratory we derive theory independent limits for both the dipole radiation and the variation of the gravitational constant. 
\end{abstract}


\bodymatter

\section{Introduction}
PSR J1012+5307 is a 5.3\,ms pulsar in a low eccentricity binary system
with orbital period of $P_b=$14.5\,h \cite{nll+95} and a low mass
helium white dwarf (WD) companion \cite{llfn95}. Ref.~\refcite{cgk98} compared the
measured optical luminosity of the WD to the value expected from WD
models and calculated a distance of $d = 840 \pm 90$\,pc. In addition
they measured, a radial velocity component of $44 \pm 8$\,km\,s$^{-1}$
relative to the solar system barycentre (SSB), and the mass ratio of
the pulsar and its companion $q=m_{p}/m_{c} = 10.5\pm 0.5$. Finally
they derived a companion mass of $m_{c} = 0.16 \pm 0.02\,M_{\odot}$, a
pulsar mass of $m_{p} = 1.64\pm 0.22\,M_{\odot}$ and an orbital
inclination angle of $i=52^\circ \pm 4^\circ$.

Ref.~\refcite{lcw+01} presented the timing analysis of PSR
J1012+5307 using 4 years of timing data from Effelsberg and 7 years
from Lovell radio telescope. They derived the spin, astrometric and
binary parameters for the system and they discussed the prospects of
future measurements of a Post-Keplerian parameter (PK) which can
contribute to the derivation of stringent limits on alternative
gravity theories.

\section{Results}
In this work PSR J1012+5307 has been revisited with seven
more years of high-precision timing data and combined
datasets from the EPTA
telescopes (Effelsberg, Lovell, Nan\c cay, Westerbork), at five different frequencies. 
The data have been analysed using the 
timing software TEMPO\footnote{http://www.atnf.csiro.au/research/pulsar/tempo/}
and all the astrometric, spin and binary 
parameters of this system have been improved.

For the first time a parallax $\pi = 1.2\pm 0.3$\,mas has been
measured for PSR J1012+5307. This corresponds to a distance of $d =
822 \pm 178$\,pc which is consistent with the $d = 840 \pm 90$ pc
measured from the optical observations. 

As predicted by Ref.~\refcite{lcw+01} a significant measurement of the change
in the orbital period of the system, $\dot{P_b}=5.0(1.4) \times
10^{-14}$, has been obtained for the first time. This is caused by the
Doppler correction (which is the combined effect of the proper motion
of the system \cite{shk70} and a correction term for the Galactic
acceleration) and by a contribution due to the quadrupole term of the
gravitational wave emission, as predicted by general relativity
(GR). After subtracting these two contributions from our measured
value, the excess value of $\dot{P}_b^{exc} = (-0.4 \pm 1.6) \times
10^{-14}$ confirms the validity of GR for one more millisecond pulsar
binary system.

All the terms mentioned above are the ones expected to contribute by
using GR as our theory of gravity. However, there are alternative
theories of gravity, that violate the strong equivalence principle
(SEP) and predict extra contributions to the observed orbital period
variation. One is the dipole term of the gravitational wave emission,
which results from the difference in gravitational binding energy of
the two bodies of a binary system. Thus PSR J1012+5307, a pulsar-WD
system, is ideal for testing the strength of such emission. For
small-eccentricity pulsar-WD systems, where the sensitivity $s$
(related to the gravitational self-energy of a body) of the WD is much
smaller than the one of the pulsar, one finds $\dot{P_{b}}^{dipole} =
-4\pi^2 \, \frac{T_\odot \mu}{P_b} \, \kappa_{D} {s_p}^{2}$,
\cite{wil01} where $T_\odot = 4.9255\,\mu s$ and $\mu$ is the reduced
mass; $s_p$ is the sensitivity of the pulsar and $\kappa_{D}$ refers
to the dipole self-gravitational contribution.

Another term is predicted by a hypothetical variation of the locally
measured gravitational constant as the universe expands,
$\dot{P}_b^{\dot{G}} = -2\,\frac{\dot{G}}{G} \left[1 -
  \left(1+\frac{m_c}{2M}\right) s_p\right] P_{b} $, \cite{dgt88,
  nor90} where $M$ is the total mass of the system. It has been shown
that there is no need to add these extra contributions to explain the
variations of the orbital period, however the excess value has been
used to set limits for a wide class of alternative theories of
gravity.

PSR J1012+5307 is an ideal lab for constraining the dipole radiation
term because the WD nature of the companion is affirmed optically, the
mass estimates are free of any explicit strong-field effects and the
mass of the pulsar is rather high, which is important in the case of
strong field effects that occur only above a certain critical mass,
like the spontaneous scalarisation \cite{de93}. Thus, by using the
$\dot{G}/G = (4 \pm 9) \times 10^{-13}$\,yr$^{-1}$ limit from the Lunar
Laser Ranging (LLR) \cite{wtb04} the $\dot{G}/G$ contribution has been
calculated and subtracted from our excess value in order to finally
obtain an improved generic limit for the dipole contribution of
$\kappa_D = (0.2 \pm 2.4) \times 10^{-3}$ (95 per cent C.L.).

A generic test for $\dot G$ cannot be done with a single binary
pulsar, since, in general, theories that predict a variation of the
gravitational constant typically also predict the existence of dipole
radiation. This degeneracy has been broken here in a joint analysis of
PSR J1012+5307 and PSR J0437$-$4715 \cite{vbv+08}, two binary
pulsar-WD systems with tight limits for ${\dot P}_b$ and different
orbital periods. By applying Eq.~1 
\begin{equation}
\frac{\dot{P}_b^{exc}}{P_b} =
- 2 \frac{\dot G}{G} \left[1 - \left(1+\frac{m_c}{2M}\right)
  s_p\right] - 4 \pi^2 \frac{T_\odot\mu}{P_b^2} \, \kappa_{\rm D}
s_p^2 
\end{equation}
 to both binary pulsars, and solving in a
Monte-Carlo simulation this set of two
equations, stringent and generic limits based purely on pulsar data
and in the strong field regime have been obtained. With a 95 per cent
C.L., $\frac{\dot G}{G} = (-0.7 \pm 3.3) \times 10^{-12} \; {\rm
  yr}^{-1}$ and $\kappa_{\rm D} = (0.3 \pm 2.5) \times 10^{-3}$.  In
the future, more accurate measurements of $\dot{P_b}$ and distance of
the two pulsars and WDs could constrain even more our derived limits.

A more detailed work on this study of
PSR J1012+5307 can be found in Ref.~\refcite{lwj+09}.

\section*{Acknowledgments}
We are very grateful to all staff at the Effelsberg, Westerbork, Jodrell Bank
and Nan\c cay radio telescopes for their help with the observations. KL was supported for this research through a stipend from the
International Max Planck Research School (IMPRS) for Astronomy and
Astrophysics at the Universities of Bonn and Cologne. MP is supported by a grant from
the Science and Technology Facilities Council (STFC). We are grateful to 
Paulo Freire for valuable discussions.

%\bibliographystyle{ws-procs975x65}
%\bibliography{ws-pro-sample}

\begin{thebibliography}{10}

\bibitem{nll+95}
L.~Nicastro, A.~G. Lyne, D.~R. Lorimer, P.~A. Harrison, M.~Bailes and B.~D.
  Skidmore, {\em MNRAS} {\bf 273}, L68 (1995).

\bibitem{llfn95}
D.~R. Lorimer, A.~G. Lyne, L.~Festin and L.~Nicastro, {\em Nat.} {\bf 376}, 393
  (1995).

\bibitem{cgk98}
P.~J. Callanan, P.~M. Garnavich and D.~Koester, {\em MNRAS} {\bf 298}, 207
  (1998).

\bibitem{lcw+01}
C.~Lange, F.~Camilo, N.~Wex, M.~Kramer, D.~Backer, A.~Lyne and O.~Doroshenko,
  {\em MNRAS} {\bf 326}, 274 (2001).

\bibitem{shk70}
I.~S. Shklovskii, {\em Soviet Ast.} {\bf 13}, 562 (1970).

\bibitem{wil01}
C.~Will, {\em Living Reviews in Relativity} {\bf 4}, 1 (2001), URL (Cited on
  2006/02/01): http://www.livingreviews.org/Irr-2001-4.

\bibitem{dgt88}
T.~Damour, G.~W. Gibbons and J.~H. Taylor, {\em Phys.~Rev.~Lett.} {\bf 61},
  1151 (1988).

\bibitem{nor90}
K.~Nordtvedt, {\em Phys.~Rev.~Lett.} {\bf 65}, 953 (1990).

\bibitem{de93}
T.~Damour and G.~Esposito-Farese, {\em Phys.~Rev.~Lett.} {\bf 70}, 2220 (1993).

\bibitem{wtb04}
J.~G. Williams, S.~G. Turyshev and D.~H. Boggs, {\em Phys.~Rev.~Lett.} {\bf
  93}, 261101 (2004).

\bibitem{vbv+08}
J.~P.~W. Verbiest, M.~Bailes, W.~van Straten and et~al., {\em ApJ} {\bf 679},
  675 (2008).

\bibitem{lwj+09}
K.~Lazaridis, N.~Wex, A.~Jessner and et~al., {\em MNRAS} {\bf 400}, 805 (2009).

\end{thebibliography}

\end{document}